\documentclass[a4paper,twocolumn,11pt,accepted=2024-11-12]{quantumarticle}
\pdfoutput=1
\usepackage[utf8]{inputenc}
\usepackage[english]{babel}
\usepackage[T1]{fontenc}
\usepackage{amsmath,amsfonts,amssymb}
\usepackage{physics}
\usepackage{graphicx}
\usepackage{dcolumn}
\usepackage{bm}
\usepackage{qcircuit}
\usepackage{hyperref}
\usepackage{tikz}
\usetikzlibrary{3d}
\newcommand{\blu}{\color{black}}

\begin{document}

\title{Construction of perfect tensors using biunimodular vectors}

\author{Suhail Ahmad Rather}
 \affiliation{Max Planck Institute for the Physics of Complex Systems, 01187 Dresden, Germany}
 \email{suhail@pks.mpg.de}
\begin{abstract}
Dual unitary gates are highly non-local two-qudit unitary gates that have been studied extensively in quantum many-body physics and quantum information in the recent past. A special class of dual unitary gates consists of rank-four perfect tensors that are equivalent to highly entangled multipartite pure states called absolutely maximally entangled (AME) states. In this work, numerical and analytical constructions of dual unitary gates and perfect tensors that are diagonal in a special maximally entangled basis are presented. The main ingredient in our construction is a phase-valued (unimodular) two-dimensional array whose discrete Fourier transform is also unimodular. We obtain perfect tensors for several local Hilbert space dimensions, particularly, in dimension six. A perfect tensor in local dimension six is equivalent to an AME state of four qudits, denoted as AME(4,6). Such a state cannot be constructed from existing constructions of AME states based on error-correcting codes and graph states. 
{\blu An explicit construction of AME(4,6) states is provided in this work using two-qudit controlled and single-qudit gates making it feasible to generate such states experimentally}.
\end{abstract}

\maketitle

\section{Introduction}
The creation and distribution of entanglement in a controlled way is a crucial requirement for performing quantum information tasks. In models of quantum computing based on quantum circuits, entanglement between different qubits, which are initially in an unentangled or product state, is created by applying entangling two-qubit gates \cite{Nielsen2010}. In currently available noisy intermediate-scale quantum (NISQ) devices \cite{preskill2018quantum}, a sequence of such two-qubit gates between different qubits results in highly entangled multiqubit states. These states are used in quantum supremacy experiments \cite{QSupreme} and are emerging test beds for realisation of unconventional phases of matter \cite{X_Mi_2021}. 

Decomposition of unitary gates into simpler ones is of importance both theoretically as well as experimentally \cite{cirac2001,KBG01,Hammerer2002,Nielsen2003}. Different entanglement measures have been defined to quantify the non-local properties of bipartite unitary gates \cite{Zanardi2000,Zanardi2001,Nielsen2003,Bhargavi2017}. These entanglement measures provide insights about the manifold of unitary gates and are useful for constructing highly entangled multipartite pure states. Analogous to bipartite pure states, a simple and elegant entanglement measure for bipartite unitary gates is the Schmidt rank defined using the Schmidt decomposition \cite{Nielsen2003,Nechita_2017}. Another important entanglement measure is entangling power which quantifies how much entanglement, on an average, a bipartite unitary gate produces when applied on uniformly distributed product states \cite{Zanardi2001}. Throughout this work we assume the dimension of local Hilbert spaces to be the same and denote it by $d$.

{\blu Under the operator-state isomorphism bipartite unitary operators that maximize the entangling power correspond to pure states of four qudits that have all two-qudit reduced density matrices maximally mixed \cite{Zanardi2000,Clarisse2005,SAA2020}}. A multipartite pure state in which all reduced density matrices are maximally mixed is called absolutely maximally entangled (AME) state \cite{Helwig_2012}. These states, by definition, have maximal entanglement in any bipartition {\blu of the system into two disjoint subsystems. AME states} have useful applications in quantum error correction \cite{Sc04,huber2018bounds}, quantum secret sharing, and quantum parallel teleportation \cite{Helwig_2012}. Existence of AME {\blu states} for general multipartite quantum systems is a difficult problem and the answer is known only in some cases \cite{Huber_AME_Table,Huber_2017,huber2018bounds}. We denote AME state of $K$ qudits each having local dimension $d$ as AME($K,d$).

A surprising result is that there is no four-qubit pure state that is an AME state i.e., AME($4,2$) {\blu state} does not exist \cite{Higuchi2000}. For qubits, the existence of AME states is completely known, AME($K$,2) exist only for $K=2,3,5,$ and $6$ \cite{Sc04,Huber_2017,huber2018bounds}. Four-party AME states; AME$(4,d)$, are equivalent to perfect tensors having four indices each running from $0$ to $d-1$ \cite{Pastawski_2015,Goyeneche2015}. Reshaped into $d^2 \times d^2\ $ matrices AME$(4, d)$ states lead to unitary matrices that remain unitary under realignment and partial transpose matrix rearrangements \cite{Zyczkowski2004}. Such bipartite unitary operators are called 2-unitary, and there exists a correspondence between AME states of even number of qudits and multi-unitary gates \cite{Goyeneche2015}. Bipartite unitary gates that remain unitary under realignment are maximally entangled unitaries and are highly non-local \cite{SAA2020}. Maximally entangled unitaries are called dual unitary gates in quantum many-body physics and have been studied extensively in the recent past \cite{Akila2016,Gopalakrishnan2019,Bertini2019,claeys2020maximum,claeys2020ergodic,ASA_2021,gutkin2020exact,suzuki2021computational,claeys2022emergent,zhou2022maximal,borsi2022remarks,singh2022ergodic,Ho_2022_Exact}. In (1+1)-dimensional discrete space-time, dual unitary gates remain unitary when the temporal and spatial axes are exchanged. This space-time duality plays a pivotal role in obtaining exact analytical results in dual unitary quantum circuits \cite{Bertini2018,bertini2019entanglement,piroli2020exact}.

Besides the non-existence of AME(4,2) state, there is a highly non-trivial result in four-party AME states concerning the local dimension $d=6$. It was shown in Ref.~\cite{Clarisse2005} that a special class of AME($4,d$) states can be obtained from permutations matrices of size $d^2$ that achieve the maximum entangling power. These ``maximally entangling" permutations are in one-to-one correspondence with Sudoku-type arrangements called orthogonal Latin squares \cite{Goyeneche2015,GRMZ_2018}. Orthogonal Latin squares exist for all $d$ except $d=2$ and $6$ \cite{bose1960further}.
As AME($4,2$) state does not exist, among the four-party AME states, $d=6$ was the only case in which the existence was not known till recently. The existence of AME($4,6$) state featured in well-known open problem lists in quantum information \cite{Open2020} and was solved recently in Ref.~\cite{SRatherAME46}. It was shown that although maximally entangling or 2-unitary permutation matrices of size $36$ do not exist, 2-unitary matrices containing complex entries exist. 

In this work, we provide a construction of AME($4,d$) states or, equivalently, perfect tensors in SU($d^2$) that are diagonal in a maximally entangled basis for bipartite state space $\mathbb{C}^d \otimes \mathbb{C}^d$. Our main interest is in local dimension six as all existing constructions of AME($4,d$) states based on error correcting codes \cite{Goyeneche2015}, combinatorial designs \cite{Goyeneche_2012_Genuinely,Goyeneche2015,GRMZ_2018}, and graph states \cite{Helwig_2013} fail in $d=6$. The maximally entangled basis in $\mathbb{C}^d \otimes \mathbb{C}^d$ used in this work is obtained from the Weyl-Heisenberg basis \cite{weylqgroup}, which is a ``nice unitary error basis'' \cite{knill1996group,AndreasMono}. It has been shown in Ref.~\cite{Tyson_2003} that a unitary operator diagonal in a maximally entangled basis is dual unitary or, equivalently, maximally entangled unitary iff the discrete Fourier transform of the phase-valued (unimodular) two-dimensional array consisting of its eigenvalues is also unimodular. The conditions for dual unitarity were also given in Ref.~\cite{yu2023hierarchical} in terms of a set of cyclic orthogonality conditions involving $d^2$ eigenvalues of the given unitary operator. Equivalently, these cyclic orthogonality conditions mean that the phase-valued $d \times d$ matrix has vanishing periodic autocorrelations. 
\begin{figure}
    \centering
\begin{tikzpicture}[scale=0.6, transform shape]
    
    \pgfmathsetmacro{\R}{2}
    \pgfmathsetmacro{\r}{0.5} 
    \draw[line width=0.5mm, gray] (0,0,0) circle (\R+\r);
    \draw[line width=0.5mm, gray] (0,0,0) circle (\R-\r);
    
    \foreach \theta in {0,90,...,360}
        \draw[line width=0.5mm,gray, dashed, domain=0:360, samples=200, smooth, variable=\phi]
            plot ({(\R+\r*cos(\phi))*cos(\theta)}, {(\R+\r*cos(\phi))*sin(\theta)}, {\r*sin(\phi)});
            
    \fill[red] (\R,-0.2,0) circle (0.12);
    \node[red, above] at (1.95,-1.1,0) {\Large{$\ket{\Lambda}$}};
    \draw [-stealth](3,0,0) -- (5,0);
    \node at (4,0.5,0) {\huge{$F_2$}};
    \end{tikzpicture}
\begin{tikzpicture}[scale=0.6, transform shape]
\pgfmathsetmacro{\R}{2}
    \pgfmathsetmacro{\r}{0.5} 
    \draw[line width=0.5mm, gray] (0,0,0) circle (\R+\r);
    \draw[line width=0.5mm, gray] (0,0,0) circle (\R-\r);
    
    \foreach \theta in {0,90,...,360}
        \draw[line width=0.5mm,gray,dashed, domain=0:360, samples=100, smooth, variable=\phi]
            plot ({(\R+\r*cos(\phi))*cos(\theta)}, {(\R+\r*cos(\phi))*sin(\theta)}, {\r*sin(\phi)});
            
    \fill[blue] (0.2,2,0) circle (0.12);
    \node[blue, above] at (0.7,1.35,0) {\large{$ \ket{\tilde{\Lambda}}$}};
\end{tikzpicture}
\caption{A unimodular vector $\Lambda$ of length $N$ lives on an $N$-dimensional torus. If $\Lambda$ is biunimodular with respect to $F_N$, then $F_N \ket{\Lambda}=:\ket{\Tilde{\Lambda}}$ is also unimodular and remains on the torus. This is illustrated in the figure for $N=2$.}
    \label{fig:torus}
\end{figure}

Sequences, vectors, and higher dimensional arrays having vanishing autocorrelations have been studied for a long time in different areas of mathematics \cite{Bjrck1990FunctionsOM,Gilbert_2010_Fourier,FUHR201586} and applied sciences \cite{Frank1962PhaseSP,CALABRO1967537}. These are called perfect sequences or arrays (depending on the context) and have useful applications in cryptography, digital signal processing, see Refs.\cite{Frank1962PhaseSP,CALABRO1967537,Schyndel1994ADW,Mow_1993_Thesis,Blake2017} and the references therein. For the purposes of this work, we only focus on two-dimensional perfect arrays or vectors consisting of phases whose discrete Fourier transform is also phase-valued (unimodular).

As we are interested in bipartite unitary gates, the two-dimensional arrays used in this work are equivalent to unimodular vectors of length $d^2$ that remain unimodular when acted by $F_d \otimes F_d$, where $F_d$ is the Fourier gate corresponding to the quantum Fourier transform on $\mathbb{C}^d$. A unimodular vector of length $N$ is said to be biunimodular if it remains unimodular when {\blu transformed} by $F_N$. A unimodular vector $\ket{\Lambda}$ of length $N$ lives on an $N$-dimensional torus and if it is biunimodular with respect to $F_N$, then $F_N\ket{\Lambda}=:\ket{\Tilde{\Lambda}}$ remains on the torus as illustrated in Fig.~(\ref{fig:torus}) for $N=2$. The term biunimodular vector is now used in a general way by replacing $F_N$ with a unitary matrix in $\mathbb{U}(N)$ \cite{FUHR201586,IDEL201576}. In quantum information biunimodular vectors have also appeared in the context of mutually unbiased bases (MUBs) \cite{Bengtsson_2007_Mutually,grassl2009sicpovms}. The biunimodular vectors used in this work are unimodular vectors of length $N=d^2$ that remain unimodular with respect to $F_d \otimes F_d$, and are equivalent to unimodular perfect arrays of size $d \times d$ .

The perfect tensors obtained in this work using biunimodular vectors can be written as a product of two-qudit controlled unitaries along with local gates as shown in Fig.~(\ref{fig:quant_cir}). Apart from generalized two-qudit {\sc cnot} gate and the Fourier gate $F_d$, the construction involves a diagonal unitary $D(\Lambda)$ obtained from the biunimodular vector $\Lambda$. The construction of these biunimodular vectors that lead to perfect tensors or, equivalently, AME($4,d$) states is discussed in this work. The quantum circuit representation shown in Fig.~(\ref{fig:quant_cir}) is useful for designing quantum circuits for AME states \cite{Alba_2019,pozsgay2023tensor}. The direction of time in all quantum circuits presented in this work runs from left to right.
\begin{figure}
    \centering
    $
    \Qcircuit @C=1em @R=4em {
       & \qw & \ctrl{1} & \qw & \gate{F_d} & \qw & \ctrl{1} & \qw & \gate{F_d} & \qw & \ctrl{1} & \qw \\
        & \qw & \targ & \qw & \qw & \qw & \gate{\Lambda} & \qw & \qw & \qw & \targ  & \qw
        }
        $
    \caption{Quantum circuit decomposition of a class of perfect tensors found in this work. Apart from the single-qudit Fourier gate, $F_d $, the decomposition involves two controlled gates: one is the generalized {\sc cnot} gate and the other one is a controlled-phase unitary obtained from biunimodular vector $\Lambda$. The construction of biunimodular vectors that lead to perfect tensors is the main focus of this work.}
    \label{fig:quant_cir}
\end{figure}

\section{Diagonal decomposition using maximally entangled basis}
The non-local properties of two-qubit unitary gates is well-understood. Any two-qubit gate $U$ in SU(4) can be written in the canonical or Cartan form as \cite{KC01,KBG01}
\begin{equation}
    U=(u_1 \otimes u_2)V(v_1 \otimes v_2),
    \label{eq:Can_form_SU4}
\end{equation}
where the $u_i$s and $v_i$s are single-qubit gates. The non-local part $V$ is given by 
\begin{equation}
    V=\exp\left[i(c_1 X \otimes X +c_2 Y \otimes Y +c_3 Z \otimes Z)\right], 
    \label{eq:Nl_part_SU4}
\end{equation}
where real numbers $c_i$s are called Cartan coefficients and $X$, $Y$ and $Z$ are the Pauli matrices. Note that only 3 parameters out of total 15 parameters appear in the non-local part. Consider a maximally entangled basis in $\mathbb{C}^2 \otimes \mathbb{C}^2$ (also called Bell basis),
\begin{equation}
\begin{split}
    \ket{\Psi_{0,1}}=\frac{1}{\sqrt{2}}\left(\ket{0}\ket{0}\pm \ket{1}\ket{1}\right), \\
    \ket{\Psi_{2,3}}=\frac{1}{\sqrt{2}}\left(\ket{0}\ket{1}\pm \ket{1}\ket{0}\right).
    \end{split}
\end{equation}
In the Bell basis $V$ is diagonal,
\begin{equation}
    V=\sum_{k=0}^3 e^{ i \mu_k} \ket{\Psi_k}\bra{\Psi_k},
\end{equation}
For bipartite unitary gates $U$ in SU($d^2$) with $d>2$, canonical decomposition such as the one given in Eq.~(\ref{eq:Can_form_SU4}) is not known. Motivated by the two-qubit case, we focus on bipartite unitary gates that are diagonal in a maximally entangled basis. Furthermore, we consider a maximally entangled basis that satisfies nice group-theoretic properties. Consider a unitary gate $U$ in SU($d^2$) having a diagonal decomposition of the following form:
\begin{equation}
    U= \sum_{a,b=0}^{d-1} \lambda_{a,b} \ket{\Phi_{ab}}\bra{\Phi_{ab}},
    \label{eq:U_Diag_Max_Ent}
\end{equation}
where the $\lambda_{a,b}$s are phases; $\lambda_{a,b} \in \text{U}(1)$, and  $\ket{\Phi_{ab}}$s are obtained from the Weyl-Heisenberg operator basis \cite{Schwinger_1960,weylqgroup} as described below. The basis operators can be obtained from the following operators or generators:
\begin{equation}
\begin{split}
    X \ket{k}=\ket{k \oplus 1},\\
    Z\ket{k}=\omega^k \ket{k},
    \label{eq:Gen_XZ}
    \end{split}
\end{equation}
where $\oplus$ is addition modulo $d$ and $\omega=\exp(2 \pi i/d)$ is root of unity. The generators satisfy the following relations:
\begin{equation}
    X^d=Z^d=\mathbb{I},\quad ZX=\omega XZ, \quad F_{\blu d}XF_{\blu d}^{\dagger}=Z,
\end{equation}
where $F_d $ is the Fourier gate whose matrix elements in the computational basis are given by
\begin{equation}
    \bra{k} F_{\blu d} \ket{l}=\omega ^{kl}{\blu /\sqrt{d}}.
\end{equation}
The maximally entangled basis $\ket{\Phi_{ab}}$ ($a,b=0,1,\cdots,d-1$) in $\mathbb{C}^d \otimes \mathbb{C}^d$ are defined using the Weyl-Heisenberg (unitary) operators: $\left\lbrace X^a Z^b|a,b=0,1,\cdots,d-1 \right\rbrace$, via the operator-state mapping \cite{Zanardi2000,Zanardi2001} as follows:
\begin{equation}
   \ket{\Phi_{ab}}:= \ket{Z^{a}X^{-b}} \equiv \frac{1}{\sqrt{d}}(Z^a X^{-b} \otimes \mathbb{I}) \ket{\Phi},
    \label{eq:Max_Ent_Bas}
\end{equation}
where $\ket{\Phi}=\sum_{m=0}^{d-1} \ket{m}\ket{m}/\sqrt{d}$ is the generalized Bell state. Note that $\ket{\Phi_{00}}=\ket{\Phi}$ and each $\ket{\Phi_{ab}}$ is the row vectorization of a given $Z^{a}X^{-b}$ (up to normalization). The ``UBB conjecture" (UBB stands for U times Bell state is Bell state) in Ref.~\cite{Shrigyan_2022} states that for a bipartite unitary operator $U$ on $\mathbb{C}^d \otimes \mathbb{C}^d$, there always exists a pair of maximally entangled states, $\ket{\Psi_1}$ and $\ket{\Psi_2}$, such that $U \ket{\Psi_1}=\ket{\Psi_2}$. Bipartite unitary operators of the form given in Eq.~(\ref{eq:U_Diag_Max_Ent}) have $d^2$ eigenvectors that are maximally entangled states, therefore these satisfy the UBB conjecture strongly by construction.

\section{Search for dual unitaries and perfect tensors}
The unitary operators discussed in this work remain unitary under the realignment and partial transpose matrix rearrangements that are defined in a product basis, respectively, as follows:
\begin{equation}
\begin{split}
    \bra{m}\bra{n} U^R\ket{k}\ket{l}:= \bra{m}\bra{k} U \ket{n}\ket{l},\\
    \bra{m}\bra{n} U^{\Gamma}\ket{k}\ket{l}:= \bra{m}\bra{l} U\ket{k}\ket{n}.
    \end{split}
\end{equation}
For $U$ acting on $\mathbb{C}^d \otimes \mathbb{C}^d$, both $U^R$ and $U^{\Gamma}$ are matrices of size $d^2$. Unitary $U$ for which $U^R$ ($U^{\Gamma}$) is also unitary is called a dual ($\Gamma$-dual) unitary. Unitary operator which is both dual and $\Gamma$-dual is called 2-unitary \cite{Goyeneche2015}. Using operator-state isomorphism \cite{Zanardi2000,Zanardi2001}, a bipartite unitary operator $U \in \mathbb{U}(d^2)$ can be used to define a pure state of four qudits each of local dimension $d$ given by
\begin{equation}
    \ket{U}=\frac{1}{d}\sum_{k,l,m,n=0}^{d-1} U_{kl}^{mn}\ket{m}\ket{n}\ket{k}\ket{l},
\end{equation}
where $U_{kl}^{mn}:=\bra{m}\bra{n} U\ket{k}\ket{l}$. Labelling the qudits from 1 to 4, the two-qudit reduced density matrices corresponding to three possible balanced bipartitions; $12/34,13/24,14/23$, are given by
\begin{equation}
    \rho_{12}=\frac{1}{d^2}UU^{\dagger},\rho_{13}=\frac{1}{d^2}U^RU^{R \dagger},\rho_{14}=\frac{1}{d^2}U^{\Gamma}U^{\Gamma \dagger}.
\end{equation}
If $U$ is 2-unitary, then $\ket{U}$ is maximally entangled in all three bipartitions ; $ \rho_{12}=\rho_{13}=\rho_{14}=\mathbb{I}/d^2$, and is therefore an AME($4,d$) state \cite{Goyeneche2015}. Equivalently, the four-index tensor $T$ defined as $T_{klmn}:=U_{kl}^{mn}$ is perfect \cite{Pastawski_2015}. A tensor is said to be perfect if any partitioning of its indices into two disjoint sets $A$ and $B$ with $|A|<|B|$ (here | . | denotes the cardinality or size of the set) defines an isometry from $A$ to $B$. {\blu In this work we use 2-unitary gate, perfect tensor and AME state interchangeably.}

The conditions of dual and $\Gamma$-dual unitarity for the unitary operator given in Eq.~(\ref{eq:U_Diag_Max_Ent}), are described below (details about the derivations can be found in appendix \ref{app:dual_T_dual}):
\begin{enumerate}
    \item Dual unitarity: The unitary of $U^R$; $U^R U^{R \dagger}= U^{R \dagger}U^R=\mathbb{I}$, leads to the following conditions for $\lambda_{a,b} \in \mathbb{U}(1)$:
    \begin{equation}
        \sum_{a,b=0}^{d-1} \lambda_{a,b} \lambda_{a+k,b+l}^*=0,\quad (k,l) \neq (0,0),
        \label{eq:lam_dual}
    \end{equation}
where $k$ and $l$ both take values in $\mathbb{Z}_d=\left\lbrace0,1,2,\cdots,d-1\right\rbrace$ and addition is defined modulo $d$. These conditions were also obtained in Ref.~\cite{yu2023hierarchical}. The above conditions mean that the phase-valued or unimodular two-dimensional array defined as
\begin{equation}
\Lambda :=\left(\begin{array}{cccc}
\lambda_{0,0} & \lambda_{0,1} & \cdots & \lambda_{0,d-1} \\
\lambda_{1,0} & \lambda_{1,1} & \cdots & \lambda_{1,d-1} \\
\vdots & \vdots &   & \vdots \\
\lambda_{d-1,0} & \lambda_{d-1,1} & \cdots  & \lambda_{d-1,d-1} \\
\end{array}
\right),
\label{eq:Lamda_def}
\end{equation}
satisfies cyclic orthogonality conditions or, equivalently, it has a vanishing autocorrelation.

One of the useful results known about unimodular vectors is that a unimodular vector has a vanishing autocorrelation iff its discrete Fourier transform (DFT) is also unimodular \cite{FUHR201586}. It was shown in Ref.~\cite{Tyson_2003} that the diagonal operator $U$ defined in Eq.~(\ref{eq:U_Diag_Max_Ent}) is dual unitary (maximally entangled) iff the discrete Fourier transform  of the unimodular two-dimensional array $\Lambda$ defined in Eq.~(\ref{eq:Lamda_def}) is also unimodular. As discussed in the appendix, this means the $d^2$-dimensional vector $\ket{\Lambda}$ (obtained from the vectorization of $\Lambda$) is biunimodular with respect to $F_d \otimes F_d$ i.e., $(F_d \otimes F_d) \ket{\Lambda}$ is also unimodular. Therefore, unimodular vectors that are biunimodular with respect to $F_d \otimes F_d$ have vanishing periodic autocorrelations given by Eq.~(\ref{eq:lam_dual}).

Constructions of biunimodular vectors with respect to $F_N$ are closely related to the cyclic-$N$ root problem \cite{Bjrck1990FunctionsOM,BJORCK1991329,Bjrck1995NewCO}. For $N=2$ and $3$, all biunimodular vectors can be obtained from Gauss sequences; $\exp\left[\frac{2 \pi i}{N} k^2\right]$ \cite{Gilbert_2010_Fourier}. For $N=4$, all biunimodular vectors are of the form \cite{Gilbert2010}
\begin{equation}
    (1,e^{i \theta},1,-e^{i \theta})\;  \text{or} \;  (1,e^{i \theta},-1,e^{i \theta}) \quad \theta \in \mathbb{R},
    \label{eq:biuni_d_2}
\end{equation}
and thus there are continuously infinite biunimodular vectors for $F_4$. When $N$ is divisible by a square, there exist continuously infinite biunimodular vectors for $F_N$ \cite{Bjrck1995NewCO}. However, our main interest is in unimodular vectors in $N=d^2$ that are biunimodular with respect to $F_d \otimes F_d$. Interestingly, in the two-qubit case $d^2=4$, we find that the above biunimodular vectors for $F_4$ are also biunimodular with respect to  $F_2 \otimes F_2=H\otimes H$, where $$\quad H=\frac{1}{\sqrt{2}} \left( \begin{array}{cc}
     1 & 1 \\
     1 & -1
\end{array}
\right),$$
is the well-known single-qubit Hadamard gate. Therefore, the one-parameter family of biunimodular vectors for $H \otimes H$ results in a one-parameter family of two-qubit dual unitaries. It is known that, up to local unitaries, two-qubit dual unitaries form a one-parameter family \cite {Bertini2019}. Thus two-qubit dual unitaries are in a one-to-one correspondence with biunimodular vectors given in Eq.~(\ref{eq:biuni_d_2}). Exhaustive parameterizations of dual unitaries are not known in larger local dimensions, and it is not clear whether such a correspondence between biunimodular vectors of length $d^2$ and dual unitary operators on $\mathbb{C}^d \otimes \mathbb{C}^d $ exists in $d>2$. In general, unimodular vectors that are biunimodular with respect to $F_{d^2}$ need not be biunimodular with respect to $F_{d} \otimes F_d$. A biunimodular vector of length $d$ can be used to obtain biunimodular vectors of length $d^2$ (which are of interest to us) due to the following result \cite{Gilbert_2010_Fourier}: if $\ket{\lambda_1}$ and $\ket{\lambda_2}$ are biunimodular vectors with respect to $F_d$, then $\ket{\lambda_1} \otimes \ket{\lambda_2}$ is biunimodular vector with respect to $F_d \otimes F_d$. Given $d^2$ number of phases $\lambda_{a,b}$, a necessary condition that these satisfy Eq.~(\ref{eq:lam_dual}) is given by 
\begin{equation}
    \left|\sum_{a,b=0}^{d-1} \lambda_{a,b}\right|^2=\sum_{a,b=0}^{d-1} |\lambda_{a,b}|^2=d^2.
\end{equation} 
This is called ``Balance theorem" \cite{Blake2017} and is useful for providing exhaustive enumeration of perfect sequences or arrays containing smaller number of distinct phases, see Refs.~\cite{Mow1993_enumerate,Mow_1993_Thesis,Blake2017}.

\item 
$\Gamma$-dual unitarity: The unitary of $U^{\Gamma}$; $U^{\Gamma} U^{\Gamma \dagger}= U^{\Gamma \dagger}U^{\Gamma}=\mathbb{I}$, leads to the following conditions for $\lambda_{a,b} \in \mathbb{U}(1)$:
\begin{equation}
        \sum_{a,b=0}^{d-1} \omega^{al-bk} \lambda_{a,b} \lambda_{a+k,b+l}^*=0, \; (k,l) \neq (0,0),
        \label{eq:lam_T_dual}
    \end{equation}
    The above condition is a weighted autocorrelation and unimodular vectors for which such  weighted autocorrelation vanish lead to $\Gamma$-dual unitary operators. To the best of our knowledge, unimodular vectors satisfying Eq.~(\ref{eq:lam_T_dual}) have not been studied in the contexts of perfect arrays or biunimodular vectors. 
\end{enumerate}

The task of constructing perfect tensors of the form given in Eq.~(\ref{eq:U_Diag_Max_Ent}) reduces to constructing $d^2$-dimensional biunimodular vectors with respect to $F_d \otimes F_d$ that satisfy additional constraints given in Eq.~(\ref{eq:lam_T_dual}). As perfect arrays lead to general dual unitary operators, those that lead to perfect tensors need to be {\em perfectly perfect} as they satisfy additional constraints. Several constructions of two-dimensional arrays that satisfy Eq.~(\ref{eq:lam_dual}) are given in Ref.~\cite{Blake2017}, and some of these constructions were also found independently in Ref.~\cite{yu2023hierarchical}. Interestingly, for odd dimensions, we found that some of these constructions also satisfy Eq.~(\ref{eq:lam_T_dual}) and thus lead to perfect tensors.
Consider $\lambda_{a,b}=\omega^{F(a,b)}$, where $F(a,b)=a^2+ab-b^2$, is quadratic in $a$ and $b$. Substituting $\lambda_{a,b}=\omega^{a^2+ab-b^2}$ in Eq.~(\ref{eq:lam_dual}), we obtain 
\begin{equation}
  \left(\sum_{a=0}^{d-1} \omega^{-(2k+l)a} \right)  \left(\sum_{b=0}^{d-1} \omega^{(k-2l)b} \right) =0.
  \label{eq:quad_ansatz_dual}
\end{equation}
From above equation, it follows that the conditions for dual unitarity are satisfied if $k=l=0$ is the {\em only} solution to the following modular equations:
\begin{equation}
     2k+l \, {\blu \equiv} \, 0 \pmod  d ,  \quad  {\blu k-2l \, \equiv} \, 0 \pmod d.
     \label{eq:mod_eq_dual}
\end{equation}

Similarly, substituting $\lambda_{a,b}=\omega^{a^2+ab-b^2}$ in Eq.~(\ref{eq:lam_T_dual}), we obtain
\begin{equation}
 \left(\sum_{a=0}^{d-1} \omega^{-2k a}\right) \left(\sum_{a=0}^{d-1} \omega^{2(k-l)b}\right) =0.
  \label{eq:quad_ansatz_T_dual}
\end{equation}
From above equation, it follows that the conditions for $\Gamma$-dual unitarity are satisfied if $k=l=0$ is the {\em only} solution to the following modular equations:
\begin{equation}
     2k\, {\blu \equiv} 0 \pmod d ,  \quad 2(k-l) \, {\blu \equiv} 0 \pmod d .
     \label{eq:mod_eq_T_dual}
\end{equation}
Solving Eq.~(\ref{eq:mod_eq_dual}), we obtain that $k=l=0$ is the only solution for all odd $d$, except for multiples of $5$. It is easy to check that for $d=5$, solutions, other than $(k,l)=(0,0)$, satisfying Eq.~(\ref{eq:mod_eq_dual}) are $(k,l)=(1,3),(2,1),(3,4),$ and $(4,2)$. Consequently, Eq.~(\ref{eq:lam_dual}) is not satisfied for these values of $k$ and $l$. One can obtain perfect tensors in odd dimensions that are multiples of $5$ by taking $F(a,b)=a^2+ab+b^2$, and unsurprisingly this choice of quadratic function does not lead to perfect tensors for odd dimensions that are multiples of 3.

In local dimension six, none of the constructions discussed in Refs.~\cite{Blake2017,yu2023hierarchical} result in biunimodular vectors for $F_6 \otimes F_6$ that also satisfy Eq.~(\ref{eq:lam_T_dual}). For example, Eq.~(\ref{eq:mod_eq_dual}) has $(k,l)=(0,0)$ as the only solution and thus dual unitarity is satisfied. However, Eq.~(\ref{eq:mod_eq_T_dual}) has solutions, other than $(k,l)=(0,0)$, given by $(k,l)=(0,3),(3,0),$ and $(3,3)$. As a consequence, Eq.~(\ref{eq:lam_T_dual}) is not satisfied for these values of $k$ and $l$. Therefore, $\lambda_{a,b}=\omega^{a^2+b^2-ab}$ does result in a unimodular vector that is biunimodular with respect to $F_6 \otimes F_6$ but it does not satisfy Eq.~(\ref{eq:lam_T_dual}). The difficulty in finding perfect tensors in local dimension six reflects in this approach also, and one needs to use numerical methods to obtain these. {\blu Below we discuss iterative procedures to obtain special biunimodular vectors that lead to perfect tensors.}

\section {Iterative procedure for biunimodular vectors \label{sec:biuni_algo}}
Before discussing the iterative procedure for unimodular {\blu vectors that are biunimodular} with respect to $F_d \otimes F_d$, it is worth to mention that unitary operators of the form given in Eq.~(\ref{eq:U_Diag_Max_Ent}) have a very special property that makes the numerical search of obtaining dual unitaries and perfect tensors easier. This special property is that their block structure determined by the non-zero entries in a product basis remains preserved under $R$ and $\Gamma$ operations. Such bipartite unitaries with an invariant block structure are ideal candidates for obtaining perfect tensors using numerical searches as was done in Ref.\cite{SRatherAME46}. {\blu For the sake of completeness we briefly review the numerical algorithms for generating dual unitary and 2-unitary gates from  Ref.\cite{SAA2020} in appendix \ref{app:algo}}. The invariance of block-structure under $R$ and $\Gamma$ operations is illustrated using the two-qubit case below. For $d=2$, Eq.~(\ref{eq:U_Diag_Max_Ent}) leads to a $4\times4$ unitary matrix of the form
\begin{equation}
    U=\left(\begin{array}{cc|cc}
        \alpha & 0 & 0 & \beta \\
         0 & \gamma & \delta & 0 \\ 
         \hline
         0 & \delta  & \gamma & 0 \\ 
         \beta  & 0 & 0 & \alpha \\
    \end{array}\right),
\end{equation}
{\blu where we have denoted the non-zero entries by $\alpha,\beta,\gamma,$ and $\delta$.} The realignment and partial transpose operations result in the following matrices
\begin{equation}
    U^R=\left(\begin{array}{cc|cc}
        \alpha & 0 & 0 & \gamma \\
         0 & \beta & \delta & 0 \\ 
         \hline
         0 & \delta  & \beta & 0 \\ 
         \gamma  & 0 & 0 & \alpha \\
    \end{array}\right), \, 
     U^{\Gamma}=\left(\begin{array}{cc|cc}
        \alpha & 0 & 0 & \delta\\
         0 & \gamma & \beta  & 0 \\ 
         \hline
         0 & \beta & \gamma & 0 \\ 
         \delta  & 0 & 0 & \alpha \\
    \end{array}\right),
\end{equation}
{\blu respectively. Note that both $U^R$ and $U^{\Gamma}$ have the same form as that of $U$ but these are not unitary in general.} As shown in appendix \ref{app:block_st}, bipartite unitaries of the form given in Eq.~(\ref{eq:U_Diag_Max_Ent}) can be written 
\begin{equation}
    U=P (F_d  \otimes \mathbb{I}) D(\Lambda) (F_d^{\dagger}  \otimes \mathbb{I}) P^T,
    \label{eq:U_cont_dec}
\end{equation}
where $$P=\sum_{i=0}^{d-1} \ket{i}\bra{i} \otimes X^{i}$$ is the generalized controlled-NOT gate with control on the first qudit and $D(\Lambda)$ is the diagonal unitary obtained from $\lambda_{a,b} \in \text{U}(1)$ as follows:
$$D(\Lambda)=\sum_{a.b=}^{d-1}\lambda_{a,b}\ket{a}\bra{a} \otimes \ket{b}\bra{b}.$$ The decomposition in Eq.~\ref{eq:U_cont_dec} is useful for quantum circuit implementations and provides insights about the construction of dual unitaries and perfect tensors. 

{\blu A Clifford two-qudit unitary gate under conjugation maps generalized Pauli matrices to themselves \cite{Gottesman_1998,Hostens_2005}. All the gates appearing in Eq.~\ref{eq:U_cont_dec} are Clifford unitaries except the diagonal unitary $D(\Lambda)$ which may or may not be Clifford in general. Most of the 2-unitary gates known in the literature are in fact Clifford gates. For example, a way to obtain perfect tensors or 2-unitary matrices in odd local dimensions is to concatenate two generalized {\sc cnot} gates \cite{Zanardi2000} as shown in Fig.~(\ref{fig:Cliff}). Such 2-unitary matrices are of the form,
\begin{equation}
  \mathcal{P}=P'P,
  \label{eq:PCliff}
\end{equation}
where $P=\sum_{i=0}^{d-1} \ket{i}\bra{i} \otimes X^{i}$ is the {\sc cnot} gate with control on the first qudit and $P'=\sum_{i=0}^{d-1}  X^{i} \otimes \ket{i}\bra{i}$ is the {\sc cnot} gate with control on the second qudit. Note that the 2-unitary $\mathcal{P}$ is a Clifford unitary by construction. The construction of 2-unitary gates discussed in this work can be viewed as a generalization of Eq.~\ref{eq:PCliff} as our construction (apart from single-qudits gates) involves concatenation of more than two controlled gates and results in  2-unitary gates that are not Clifford.

\begin{figure}
    \centering
    $
    \Qcircuit @C=1em @R=4em {
        & \qw & \ctrl{1} &  \qw & \targ & \qw \\
        & \qw & \targ &  \qw & \ctrl{-1} & \qw 
        }
        $
    \caption{Quantum circuit representation of a Clifford 2-unitary in odd local dimensions obtained from two generalized {\sc cnot} gates.}
    \label{fig:Cliff}
\end{figure}
}

As biunimodular vectors are the main ingredients for constructing dual unitaries and perfect tensors, we present a simple iterative procedure for obtaining biunimodular vectors from random unimodular vectors. The algorithm involves the following steps:
\begin{enumerate}
    \item Start with a unimodular vector $\Lambda^0$ of length $d^2$ consisting of random phases.
    \item Consider the action of $F_d \otimes F_d$ on $\Lambda^0$; $$\Tilde{\Lambda}^{0} :=(F_d \otimes F_d) \Lambda^0.$$ In general, $\Tilde {\Lambda}^{0}$ is not phase-valued.
    \item Ignoring the absolute values and taking only the phases of each entry in $\Tilde {\Lambda}^{0}$, we obtain a unimodular vector denoted as $\Lambda^1$ \footnote{$\Lambda^1$ is the closest unimodular vector to  $\Tilde {\Lambda}^{0}$. Author thanks Arul Lakshminarayan for pointing this out.}.
\end{enumerate}
Repeating the above procedure for $\Lambda^1$, after $n$ iterations, the algorithm converges to a unimodular vector $\Lambda^n$ such that $\Tilde {\Lambda}^{n+1}=(F_d \otimes F_d) \Lambda^n$ is also unimodular or, equivalently, $\Lambda^n$ is a biunimodular vector (up to a given precision). A similar algorithm is presented in Ref.~\cite{FUHR201586} that involves an additional step involving $F_d^{\dagger}$ which is useful for showing convergence. In the two-qubit case, up to a phase, the biunimodular vectors obtained from the algorithm are of the form $(1,e^{i \theta},1,-e^{i \theta})$ or $(1,e^{i \theta},-1,e^{i \theta})$, which are the only possible cases as mentioned above. Substituting the biunimodular vectors obtained from the algorithm in the diagonal decomposition given in Eq.~(\ref{eq:U_Diag_Max_Ent}) results in dual unitary gates.

\section {Biunimodular vectors that lead to perfect tensors}
We recall that the set of 2-unitaries is a subset of dual unitaries that satisfy additional constraint of $\Gamma$-duality (i.e., unitarity with respect to partial transpose). For smaller dimensions $d=3,4$ and $5$, the iterative procedure initiated with random phase-valued  vectors described in the previous section results in biunimodular vectors that lead to 2-unitaries. However, for larger dimensions, particularly in $d=6$, numerical results discussed in Appendix \ref{app:numerics} indicate that the probability of converging to such special biunimodular vectors is almost zero. Therefore, we slightly modify the above procedure to obtain biunimodular vectors that correspond to 2-unitaries. This is done by taking into account a simple known fact that a dual unitary $U$ is 2-unitary iff $SU$ and $US$ are also dual-unitary \cite{Zanardi2001,SAA2020}, where $S$ is the {\sc swap} gate. For general $d$, the {\sc swap} operator is not diagonal in the maximally entangled basis, $S\ket{Z^a X^{-b}}=\ket{(Z^a X^{-b})^T}$. As $\left\lbrace{Z^a X^{-b}:a,b=0,1,\cdots,d-1}\right\rbrace$ form an orthonormal basis for operators acting on $\mathbb{C}^{d}\otimes \mathbb{C}^d$,
\begin{equation}
S= \sum_{a,b=0}^{d-1}\ket{(Z^a X^{-b})^T}\bra{Z^a X^{-b}}.
\end{equation}
The above expansion provides a Schmidt decomposition of the {\sc swap} gate. 

\begin{figure}
    \centering
    \includegraphics[width=1\linewidth]{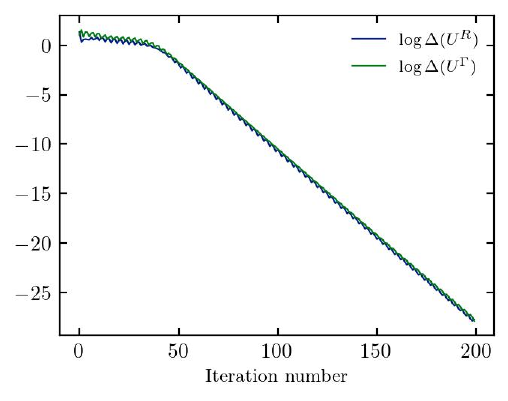}
    \caption{Convergence of the modified iterative algorithm is shown for local dimension $d=4$. Starting from a random phase-valued vector of length 16 the convergence of the algorithm to a desired biunimodular vector is shown using the log-linear plot of differences $\Delta(U^R)$ and $\Delta(U^\Gamma)$ versus the iteration number.}
    \label{fig:biuni_conv}
\end{figure}

{\blu Let $\mu_S$ denotes the unimodular vector whose entries are obtained from expansion of the {\sc swap} gate in the maximally entangled basis. We modify the iterative procedure described in the previous section by multiplying the entries of the phased-valued vectors with the corresponding entries of $\mu_S$ at each iteration. Due to this small modification, the probability of converging to perfect tensors increases appreciably, even in larger dimensions. The convergence of this modified algorithm is shown in Fig.~\ref{fig:biuni_conv} for a single random realization in local dimension $d=4$. The algorithm is initiated with a random phase-valued vector of length 16. The unitary operator $U$ is obtained from the random phase-valued vector using Eq.~\ref{eq:U_Diag_Max_Ent}. The convergence of the algorithm is studied using the following difference quantities,
\begin{align}
\Delta(U^{R}):=||U^{R}U^{R \dagger}-\mathbb{I}||,\\
\Delta(U^{\Gamma}):=||U^{\Gamma}U^{\Gamma \dagger}-\mathbb{I}||,
\end{align}
where || · · · || is the Frobenius norm. For a desired biunimodular vector the corresponding unitary operator $U$ is a perfect tensor or 2-unitary satisfying the following simple equation,
\begin{equation}
    \Delta(U^R)=\Delta(U^\Gamma)=0.
\end{equation}
As shown in Fig.~(\ref{fig:biuni_conv}) both  $\Delta(U^R)$ and $\Delta(U^\Gamma)$ approach to zero in an exponential manner. The rate of convergence depends on the initial phase-valued vector.
}

\subsection{Biunimodular vectors in local dimension six}
{\blu The modified algorithm discussed above results in special biunimodular vectors that lead to perfect tensors in local dimension $d=6$. These biunimodular vectors of length $d^2=36$ may contain many distinct phases. However}, we focus on simpler ones such as those defined over roots of {\blu unity} or over a cyclic group. For local dimension $d=6$, following biunimodular vectors of length $d^2=36$ were obtained (after removing an overall phase):

\begin{widetext}
\begin{equation}
\begin{split}
\Lambda_{1}=\exp{\frac{2\pi i}{6}\left[0 , 1 , 0 , 1 , 3 , 3 , 3 , 3 , 1 , 5 , 2 , 4 , 2 , 1 , 3 , 1 , 2 ,
       3 , 1 , 1 , 2 , 0 , 3 , 5 , 5 , 3 , 2 , 3 , 2 , 5 , 4 , 4 , 1 , 5 ,
       5 , 1 \right]},\\
       \Lambda_2=\exp{\frac{ 2 \pi i}{6} \left[0, 2, 3, 3, 2, 0, 0, 3, 2, 2, 0, 4, 2, 0, 3, 5, 0, 0, 0, 5, 0, 0,
       2, 0, 2, 2, 5, 3, 2, 4, 2, 3, 0, 2, 0, 0\right]},\\
\Lambda_{3}=\exp{\frac{2 \pi i}{3}\left[0, 2, 2, 0, 0, 1, 0, 1, 1, 1, 2, 1, 0, 2, 0, 2, 2, 2, 2, 0, 2, 2,
       2, 1, 1, 1, 2, 0, 2, 2, 0, 1, 2, 2, 1, 0 \right]}.
\end{split}
\label{eq:biuni_Lam}
\end{equation}
\end{widetext}
{\blu Denoting sixth-root of unity as $\omega$, the elements of the biunimodular vector $\Lambda_1$ are as follows: $(\Lambda_{1})_{0,0}=\omega^0,(\Lambda_{1})_{0,1}=\omega^1,\cdots,(\Lambda_{1})_{5,4}=\omega^5,(\Lambda_{1})_{5,5}=\omega^1$. Note that all sixth roots of unity appear in $\Lambda_1$, while only five and three roots appear in $\Lambda_2$ and $\Lambda_3$, respectively. It can be checked by a direct computation that above three phase-valued vectors satisfy both Eq.~\ref{eq:lam_dual} and Eq,~\ref{eq:lam_T_dual}. Substituting the above biunimodular vectors in Eq.~(\ref{eq:U_Diag_Max_Ent}), results in perfect tensors or, equivalently, 2-unitary matrices of size 36.}

The elements of the above biunimodular vectors are of the form $\Lambda_{a,b}=\omega^{\blu I(a,b)}$, where ${\blu I(a,b)}$ is an integer function. It was tempting to try quadratic and higher order polynomials in $a$ and $b$ defined from $\mathbb{Z}_6 \times \mathbb{Z}_6$ to $\mathbb{Z}_6$ that result in integer functions of the biunimodular vectors given in Eq.~(\ref{eq:biuni_Lam}). Unfortunately, we did not find polynomial functions that result in integer functions corresponding to the above biunimodular vectors. We also tried several integer functions for two-dimensional perfect arrays given in Ref.\cite{Blake2017}, but these also did not result in desired integer functions. It is quite remarkable that the iterative procedure initiated with random unimodular vectors leads to such biunimodular vectors that contain only few distinct phases. Numerically we also observed biunimodular vectors containing more than six phases, and a complete classification of these is left for future studies.

\subsection{AME(4,6) states and 2-unitary gates obtained from biunimodular vectors}
We denote a 2-unitary matrix obtained from a biunimodular vector $\Lambda$ as $\mathcal{U}(\Lambda)$. As perfect tensors are equivalent to four-party absolutely maximally entangled (AME) states, $\mathcal{U}(\Lambda)$ is equivalent to an absolutely maximally entangled state of four qudits each having local dimension $d=6$. In computational basis, the four-party AME state corresponding to $\mathcal{U}(\Lambda)$ is given by,
\begin{equation}
    \ket{\mathcal{U}(\Lambda)}=\frac{1}{6}\sum_{k,l,m,n=0}^5 \left(\mathcal{U}(\Lambda)\right)^{kl}_{mn} \ket{k}\ket{l}\ket{m}\ket{n},
    \end{equation}
where $\left(\mathcal{U}(\Lambda) \right)^{kl}_{mn}:=\bra{k}\bra{l} \mathcal{U}(\Lambda) \ket{m} \ket{n}$. 
Substituting $\Lambda=\Lambda_1,\Lambda_2$, and $\Lambda_3$ results in different AME states. {\blu Local unitary (LU) equivalence is a useful notion to classify bipartite unitary operators. A unitary operator $U$ acting on $\mathbb{C}^d \otimes \mathbb{C}^d$ is LU equivalent to $U'$ if there exist single-qudit unitary gates $u_1,u_2,v_1,$ and $v_2$ such that the following equation holds:
\begin{equation}
    U'=(v_1\otimes v_2)U(u_1 \otimes u_2).
\end{equation}
The LU equivalence of 2-unitaries or, equivalently, four-party AME states was examined in Ref.~\cite{SRVA_2022} using invariants of LU equivalence. Using the invariants that distinguish 2-unitary gates belonging to different LU classes presented in Ref.~\cite{SRVA_2022}, it can be shown that $\mathcal{U}(\Lambda_1),\mathcal{U}(\Lambda_2),$ and $\mathcal{U}(\Lambda_3)$ are not LU equivalent.}

The biunimodular vectors given in Eq.~(\ref{eq:biuni_Lam}) treated as pure state in $\mathbb{C}^6 \otimes \mathbb{C}^6$ are not factorizable i.e., these cannot be obtained from a tensor product of biunimodular vectors of length 6. For several odd dimensions, numerically we observed that initiating the algorithm with factorizable unimodular vectors results in factorizable biunimodular vectors that lead to perfect tensors. However, in even dimensions such as $d=4$ and $d=6$, we did not obtain factorizable biunimodular vectors that lead to perfect tensors. In $N=6$, there are $48$ biunimodular vectors \cite{FUHR201586} and all of them are listed in Ref.~\cite{grassl2009sicpovms}. We checked numerically all $48^2$ biunimodular vectors of length $36$ of the form $\ket{\lambda_1} \otimes \ket{\lambda_2}$, where $\ket{\lambda_{1,2}}$ are biunimodular vectors of length $6$, but none of them led to a perfect tensor in dimension six.

In computational basis, the block structure of $\mathcal{U}(\Lambda)$ determined by its non-zero entries consists of $6 \times 6$ unitary matrices as shown in Fig.~(\ref{fig:U_36_block}). Using Eq.~(\ref{eq:U_cont_dec}), $\mathcal{U}(\Lambda)$ can be written as
\begin{equation}
    \mathcal{U}(\Lambda)={\text{\sc cnot}} (F_6 \otimes  \mathbb{I}) D(\Lambda)(F_6^{\dagger}  \otimes \mathbb{I}) ({\text{\sc cnot}})^T,
\end{equation}
where {\sc cnot} is the generalized controlled-NOT gate and $D(\Lambda)$ is a diagonal unitary obtained from biunimodular vector $\Lambda$. 

From the above decomposition, we identify perfect tensors having a symmetric form given by
\begin{equation}
    \mathcal{U}'(\Lambda) ={\text{\sc cnot}} (F_6 \otimes  \mathbb{I}) D(\Lambda)(F_6  \otimes \mathbb{I}) {\text{\sc cnot}}.
    \label{eq:U_Lam_perfect}
\end{equation}
The quantum circuit representation of these perfect tensors is shown in Fig.~(\ref{fig:quant_cir}). The price one pays to obtain such symmetric form is that $\mathcal{U}'(\Lambda)$ is no longer diagonal in the maximally entangled basis considered in this work.

{\blu The absolute values of the non-zero entries of the perfect tensors or, equivalently, 2-unitary matrices of size 36 is shown in Fig.~(\ref{fig:U_36_block}). As discussed in the previous section, in general, the block structure of bipartite unitaries of the form given in Eq.~(\ref{eq:U_Diag_Max_Ent}) consists of $d$ number of $d \times d$ circulant unitary matrices. This is shown in Fig.~(\ref{fig:U_36_block}) for the perfect tensors in local dimension six, whose corresponding $6 \times 6$ unitary matrices are also shown separately. For the perfect tensor obtained from $\Lambda_1$, there are seven non-zero distinct absolute values: $1/6,1/(2\sqrt{3}),1/3,\sqrt{7}/6,1/\sqrt{3},\sqrt{13}/6,$ and $2/3$. Interestingly, for the perfect tensors obtained from the biunimodular vectors $\Lambda_2$ and $\Lambda_3$, there are only three non-zero distinct absolute values: $1/(2\sqrt{3}),1/2,$ and $1/\sqrt{3}$. Unlike the 2-unitary gates discussed in Refs.~\cite{AME46_conf,SRVA_2022} that have a block-structure consisting of nine unitary matrices each of size $4$, the perfect tensors obtained in this work have block structure consisting of six unitary matrices each of size $6$. 
}

\subsection{2-unitary complex Hadamard matrices (CHMs) obtained from biunimodular vectors}
A unifying notion for controlled unitaries, complex Hadamard matrices (Fourier gate is an example of a complex Hadamard matrix), and dual unitaries is that of biunitaries \cite{Reutter_2019,claeys2023dualunitary}. Therefore, Eq.~(\ref{eq:U_Lam_perfect}) can be viewed as a composition of biunitaries to obtain perfect tensors. Interestingly, multiplying the perfect tensors obtained from $\Lambda_{1,2,3}$ with local gates of the form $\mathbb{I} \otimes F_6$ or $ F_6 \otimes \mathbb{I}$ on both sides, results in complex Hadamard matrices of size 36, There are constructions of complex Hadamard matrices that are dual unitary \cite{gutkin2020exact,borsi2022remarks,claeys2022emergent} and 2-unitary \cite{ASA_2021,bruzda2023multiunitary}. To the best of our knowledge, none of these constructions result in 2-unitary complex Hadamard matrices of size 36. This work provides an explicit construction of 2-unitary complex Hadamard matrices (CHM) of size 36 of the following form:
\begin{equation}
    \mathcal{U}_{\text{CHM}}(\Lambda)=(F_6 \otimes \mathbb{I})  \mathcal{U}(\Lambda) (F_6^{\dagger} \otimes \mathbb{I}).
\end{equation}
Substituting $\Lambda=\Lambda_{1},\Lambda_{2},$ and $\Lambda_{3}$ given  in Eq.~(\ref{eq:biuni_Lam}) in the above equation, one obtains CHMs of size 36 that are 2-unitary. {\blu A family of 2-unitary complex Hadamard matrices obtained from our construction are of the following form:
\begin{equation}
    \mathcal{V}_{\text{CHM}}(\Lambda)=\text{\sc cz}^{\dagger} (F_6 \otimes F_6)D(\Lambda)(F_6 \otimes F_6)\text{\sc cz}
    \label{eq:UCHM}
\end{equation}
where {\sc cz} is the controlled-Z gate and $\Lambda$ is phase-valued vector satisfying both Eq.~\ref{eq:lam_dual} and Eq,~\ref{eq:lam_T_dual}. The quantum circuit representation of these 2-unitary CHMs is shown in Fig.~(\ref{fig:quant_cir_CHM}). Numerical results suggest that Eq.~\ref{eq:UCHM} results in dual or 2-unitary CHMs for general local dimension $d$. A formal proof of this claim and other properties concerning these dual and 2-unitary CHMs is  left for future studies.
\begin{figure}
    \centering
    $
    \Qcircuit @C=1em @R=4em {
        & \qw & \ctrl{1} & \qw & \gate{F_6} & \qw & \ctrl{1} & \qw & \gate{F_6} &
        \qw & \ctrl{1} & \qw \\
       & \qw & \gate{Z} & \qw & \gate{F_6} & \qw & \gate{\Lambda} & \qw &  \gate{F_6} & 
        \qw & \gate{Z^{\dagger}} & \qw
        }
        $
    \caption{Quantum circuit decomposition of a 2-unitary CHM obtained from biunimodular vector $\Lambda$. The decomposition contains single-qudit Fourier gate $F_6$ and controlled-Z gate.}
    \label{fig:quant_cir_CHM}
\end{figure}
}
\section{Summary and Future directions}
This work provides a construction for dual unitaries and perfect tensors, using unimodular two-dimensional perfect arrays or biunimodular vectors. The existing methods for constructing perfect tensors based on error correcting codes and combinatorial arrangements fail for local dimension six. In this work, explicit constructions of perfect tensors in local dimension six are given that have simple quantum circuit representations. This work opens new directions and insights to the construction of dual unitaries and perfect tensors. One of the future goals is to classify all biunimodular vectors that lead to dual unitaries, at least, for smaller local dimensions such as $d=3$ and $4$. There is correspondence between perfect binary sequences and arrays (these have $\pm 1$ as entries) and difference sets in an Abelian group \cite{JEDWAB1994241,Menon1962Difference}. One of the future goals is to study whether such correspondence exists between unimodular two-dimensional perfect arrays discussed in this work and difference sets. For larger local dimensions, there are unitary error bases (UEBs) that are not equivalent \cite{AndreasMono}. It may be interesting to search for dual unitaries and perfect tensors that are diagonal in maximally entangled bases obtained from inequivalent UEBs.

\section*{Acknowledgments}
Author is grateful to Arul Lakshminarayan and Pieter W. Claeys for discussions and detailed feedback on the manuscript. Author specially thanks Andrew Z. Tirkel and Samuel T. Blake for discussions on perfect arrays, and Michael Rampp, Jiangtian Yao, Philippe Suchsland, and Gabriel Oliveira Alves for stimulating discussions. Special thanks to Wojciech Bruzda and Karol {\.Z}yczkowski for discussions and collaborations on related works. Quantum circuit diagrams used in this work were produced using Q-circuit package \cite{eastin2004qcircuittutorial}.

\bibliographystyle{quantum}
\bibliography{ent_local}
\newpage

\onecolumngrid
\appendix
\section{Derivation of cyclic orthogonality conditions \label{app:dual_T_dual}}
In this section, we derive the cyclic orthogonality conditions; Eqs.~(\ref{eq:lam_dual}) and (\ref{eq:lam_T_dual}), arising from dual and $\Gamma$-dual unitarity, respectively. The conditions for dual unitarity of the diagonal operator defined in Eq.~(\ref{eq:U_Diag_Max_Ent}) are given in Ref.~\cite{Tyson_2003} using biunimodular functions, and recently the cyclic orthogonality conditions given in Eq.~(\ref{eq:lam_dual}) were obtained in Ref.~\cite{yu2023hierarchical}. For the sake of completeness, we reproduce them here and make the connection to biunimodular vectors as described in the main text.

Before deriving the conditions for dual and $\Gamma$-dual unitarity, we first write the Schmidt decomposition of the diagonal operator obtained in Ref.~\cite{Tyson_2003},
\begin{equation}
\begin{split}
    U & =\frac{1}{d}\sum_{a,b=0}^{d-1}\Tilde{\lambda}_{a,b} X^a Z^b \otimes (X^a Z^b)^*,\\
    & = \sum_{a,b=0}^{d-1}|\Tilde{\lambda}_{a,b}| \left(\frac{\Tilde{\lambda}_{a,b}}{|\Tilde{\lambda}_{a,b}|}\frac{1}{\sqrt{d}}X^a Z^b\right) \otimes \left(\frac{1}{\sqrt{d}}(X^a Z^b)^*\right),
    \end{split}
\end{equation}
where $\Tilde{\lambda}_{a,b}$ is the 2-dimensional (2D) discrete Fourier transform defined as
\begin{equation}
\Tilde{\lambda}_{a,b}=\frac{1}{d}\sum_{m,n=0}^{d-1} \omega^{am+bn}\lambda_{m,n}.
\end{equation}
Note that the absolute values of $\Tilde{\lambda}_{a,b}$ determine the Schmidt coefficients of $U$ and hence $U$ is maximally entangled or dual unitary iff $|\Tilde{\lambda}_{a,b}|=1$ for all $a,b=0,1,\cdots,d-1$.

Defining $d^2$-dimensional vectors $\ket{\Lambda}$ and $\ket{\Tilde{\Lambda}}$ as $\lambda_{a,b}=(\bra{a}\bra{b})\ket{\Lambda}$ and $\Tilde{\lambda}_{a,b}=(\bra{a}\bra{b})\ket{\Tilde{\Lambda}}$, respectively,  the above 2D discrete Fourier transform can be written in a single matrix equation as 
\begin{equation}
   \ket{\Tilde{\Lambda}}=(F_d \otimes F_d)  \ket{\Lambda},
\end{equation}
where $F_d $ is the $d$-dimensional Fourier gate. It follows from the row-vectorization identity; $\ket{ABC}=(A \otimes C^T)\ket{B}$, that $d\times d$ matrices $\Lambda$ and $\tilde \Lambda$ are related as $$\tilde \Lambda=F_d \Lambda F_d^T=F_d \Lambda F_d.$$ 

It is important to mention that $(F_d \otimes F_d)$ is a local gate and is different from the usual discrete Fourier transform on $\mathbb{C}^{d^2}$ denoted as $\mathcal{F}_{d^2}$. The matrix elements of $\mathcal{F}_{d^2}$ in a product basis are given by $$\bra{m}\bra{n} \mathcal{F} \ket{p}\ket{q}=\frac{1}{d} \exp\left[\frac{2\pi i}{d^2} (md+n)(pd+q)\right],$$ where $m,n,p,q$ all take values from $0$ to $d-1$. It is easy to check that $\mathcal{F}_{d^2}$ is a non-local gates unlike $(F_d \otimes F_d)$ and is in fact maximally entangled i.e., dual unitary \cite{Tyson2003}.
\subsection{Dual unitarity of the diagonal operator}
 The diagonal operator given in Eq.~(\ref{eq:U_Diag_Max_Ent}) can be written as 
\begin{equation}
        U=\frac{1}{d} \sum_{a,b=0}^{d-1} \lambda_{a,b} \ket{Z^a X^{-b}} \bra{Z^a X^{-b}}.
\end{equation}
For operators $A$ and $B$ acting on $\mathbb{C}^{d}$, $(A\otimes B)^R=\ket{A}\bra{B^*}$, where $^*$ denotes the complex conjugation. Using this identity, the realignment of $U$ is given by
\begin{equation}
        U^R=\frac{1}{d} \sum_{a,b=0}^{d-1} \lambda_{a,b} Z^a X^{-b} \otimes (Z^a X^{-b})^*.
        \label{eq:U_diag_R}
\end{equation}
Unitarity of $U^RU^{R\dagger}=\mathbb{I}$ leads to following equations:
\begin{equation}
    \begin{split}
        \frac{1}{d} \left(\sum_{a,b=0}^{d-1} \lambda_{a,b} Z^a X^{-b} \otimes (Z^a X^{-b})^*\right)
       \frac{1}{d} \left( \sum_{a',b'=0}^{d-1} \lambda_{a',b'}^* (Z^{a'} X^{-b'})^{\dagger} \otimes (Z^{a'} X^{-b'})^T\right)=\mathbb{I},\\
      \sum_{a,b=0}^{d-1} \sum_{a',b'=0}^{d-1} \lambda_{a,b} \lambda_{a',b'}^* Z^a X^{-b} (Z^{a'} X^{-b'})^{\dagger} \otimes (Z^a X^{-b})^* (Z^{a'} X^{-b'})^T =d^2 \mathbb{I},
    \end{split}
\end{equation}
The above equation can be be simplified thanks to the following nice relations satisfied by the generators:
\begin{equation}
    \begin{split}
    (X^{a}Z^{b})^{*}=X^{a}Z^{-b}, (X^{a}Z^{b})^{T}=\omega^{-ab}X^{-a}Z^{b},
    X^{a}Z^{b}X^{a'}Z^{b'}=\omega^{a'b}X^{a+a'}Z^{b+b'}
    \end{split}
    \label{eq:XZ_Com}
\end{equation}
Using these relations, the equation for unitarity of $U^R$ simplifies to the following equation:
\begin{equation}
    \begin{split}
        \sum_{a,b=0}^{d-1}\sum_{a',b'=0}^{d-1} \lambda_{a,b} \lambda_{a',b'}^* (Z^{a-a'} X^{-b+b'}) \otimes (Z^{a-a'} X^{b-b'}) =d^2\mathbb{I},
    \end{split}
\end{equation}
As the $X^aZ^b$s ($a$ and $b$ each take values from 0 to $d-1$) form an orthonormal basis for $d \times d$ matrices, comparing the coefficients on both sides we obtain the following conditions on the phases,
\begin{equation}
    \begin{split}
        \sum_{a,b=0}^{d-1} \lambda_{a,b} \lambda_{a,b}^*=\sum_{a,b=0}^{d-1}\underbrace{|\lambda_{a,b}|^2}_{=1}=d^2,\\
        \sum_{a,b=0}^{d-1}\lambda_{a,b} \lambda_{a+k,b+l}^*=0 \quad (k,l)\neq (0,0).
    \end{split}
\end{equation}
\subsection{$\Gamma$-dual unitarity of the diagonal operator}
For writing the conditions on unitarity of $U^{\Gamma}$, it is convenient to work with $U^{R\Gamma}:=\left(U^R\right)^\Gamma$. As $\left(U^R\right)^\Gamma=U^{\Gamma}S$ where $S$ is the {\sc swap} gate, unitarity of $U^{R \Gamma}$ implies unitarity of $U^{\Gamma}$.
From Eq.~\ref{eq:U_diag_R} it follows that
\begin{equation}
    U^{R \Gamma}=\frac{1}{d} \sum_{a,b=0}^{d-1} \lambda_{a,b} Z^a X^{-b} \otimes (Z^a X^{-b})^{\dagger}.
        \label{eq:U_diag_GR}
\end{equation}
The unitarity of $U^{R \Gamma}$; $U^{R \Gamma}U^{R \Gamma\dagger}=\mathbb{I}$ leads to the following equations
\begin{equation}
    \begin{split}
        \frac{1}{d} \left(\sum_{a,b=0}^{d-1} \lambda_{a,b} Z^a X^{-b} \otimes (Z^a X^{-b})^{\dagger}\right)
       \frac{1}{d} \left( \sum_{a',b'=0}^{d-1} \lambda_{a',b'}^* (Z^{a'} X^{-b'})^{\dagger} \otimes (Z^{a'} X^{-b'}) \right)=\mathbb{I},\\
       \sum_{a,b=0}^{d-1} \sum_{a',b'=0}^{d-1} \lambda_{a,b}\lambda_{a',b'}^* Z^a X^{-b} (Z^{a'} X^{-b'})^{\dagger} \otimes (Z^a X^{-b})^{\dagger} (Z^{a'} X^{-b'})=d^2\mathbb{I}.
    \end{split}
\end{equation}
Using Eq.~(\ref{eq:XZ_Com}), the above equation simplifies to the following equation
\begin{equation}
    \sum_{a,b=0}^{d-1}\sum_{a',b'=0}^{d-1} \omega^{ab+a'b'-2a'b} \lambda_{a,b} \lambda_{a',b'}^* (Z^{a-a'} X^{b-b'}) \otimes (Z^{-a+a'} X^{-b+b'}) =d^2\mathbb{I}.
\end{equation}
Proceeding as in the dual unitarity case, comparing the coefficients on both sides we obtain the following set of conditions on the phases:
\begin{equation}
    \begin{split}
        \sum_{a,b=0}^{d-1} \omega^{al-bk} \lambda_{a,b} \lambda_{a+k,b+l}^*=0 \quad (k,l)\neq (0,0).
    \end{split}
\end{equation}

\section{Block structure of diagonal unitaries in the computational basis \label{app:block_st}}

The block structure of the diagonal operator $U$ determined by its non-zero entries in the computational basis consists of $d$ unitary matrices each of size $d$ as explained below. 

Expanding $U$ in computational basis, we obtain:
\begin{equation}
\begin{split}
    U & =\frac{1}{d}\sum_{a,b=0}^{d-1} \lambda_{a,b} \left(\sum_{k,l=0}^{d-1} (Z^a X^{-b})_{kl} \ket{k} \ket{l}\right)
    \left(\sum_{k',l'=0}^{d-1} (Z^a X^{-b})_{k'l'}^* \ket{k'} \ket{l'}\right), \\
    & =\frac{1}{d}\sum_{a,b=0}^{d-1}  \lambda_{a,b} \sum_{k,l=0}^{d-1} \sum_{k',l'=0}^{d-1}
(\bra{k} Z^a X^{-b}\ket{l}) \left(\bra{k'} (Z^a X^{-b})^* \ket{l'}\right)  \ket{k} \ket{l} \bra{k'}\bra{l'}.
\end{split}  
\end{equation}
Using Eq.~(\ref{eq:Gen_XZ}), the above equation simplifies as follows:
\begin{equation}
\begin{split}
    U & = \frac{1}{d}\sum_{a,b=0}^{d-1}  \lambda_{a,b} \sum_{k,l=0}^{d-1} \sum_{k',l'=0}^{d-1}
\omega^{a (k-k')} \underbrace{\langle{k}|l-b\rangle}_{\delta_{k,l-b}}  \underbrace{\langle{k'}|l'-b\rangle}_{\delta_{k',l'-b}}  \ket{k} \bra{k'} \otimes \ket{l} \bra{l'}, \\
& =\frac{1}{d}\sum_{a,b=0}^{d-1}  \lambda_{a,b} \sum_{k,k'=0}^{d-1}
\omega^{a (k-k')} \ket{k} \bra{k'} \otimes \ket{k+b} \bra{k'+b}.\\
& =\frac{1}{d}\sum_{a,b=0}^{d-1}  \lambda_{a,b} \sum_{k,k'=0}^{d-1}
\omega^{a (k-k')} \ket{k} \bra{k'} \otimes X^k \ket{b} \bra{b} (X^T)^{k'}.
\end{split}  
\end{equation}
Multiplying $U$ by the generalized {\sc cnot} gate; $P=\sum_{m=0}^{d-1} {\blu \ket{m} \bra{m}} \otimes X^m$, on the  right and $P^T$ on the left, we obtain

\begin{equation}
\begin{split}
    P^TUP & =\frac{1}{d}\sum_{a,b=0}^{d-1}  \lambda_{a,b} \sum_{m,n=0}^{d-1} \sum_{k,k'=0}^{d-1}
\omega^{a (k-k')} \underbrace{\langle{m}\ket{k}}_{\delta_{m,k}} \underbrace{\langle{k'}\ket{n}}_{\delta_{k',n}} \ket{m} \bra{n} \otimes (X^T)^m X^k \ket{b} \bra{b} (X^T)^{k'} X^n,\\
& =\frac{1}{d}\sum_{a,b=0}^{d-1}  \lambda_{a,b} \sum_{k,k'=0}^{d-1}
\omega^{a (k-k')} \ket{k} \bra{k'} \otimes \ket{b} \bra{b}.
\end{split}  
\end{equation}
Note that $P^TUP$ is a controlled unitary; second qudit acts as a control and first qudit is the target, and each unitary acting on the first qudit is a circulant matrix. The above equation simplifies to 
\begin{equation}
    P^TUP=(F_d \otimes \mathbb{I}) \left(\sum_{a,b=0}^{d-1} \lambda_{a,b} \ket{a}\bra{a} \otimes  \ket{b}\bra{b}\right)(F_d^{\dagger} \otimes \mathbb{I}),
\end{equation}
where $F_d$ is the Fourier gate. Note that $D(\Lambda):=\sum_{a,b=0}^{d-1} \lambda_{a,b} \ket{a}\bra{a} \otimes  \ket{b}\bra{b}$, is diagonal unitary and is also a controlled unitary. Therefore, we obtain a very useful decomposition for bipartite unitaries of the form given in Eq.~(\ref{eq:U_Diag_Max_Ent}) in terms of controlled unitaries,
\begin{equation}
    U=P(F_d\otimes \mathbb{I}) D(\Lambda) (F_d^{\dagger}  \otimes \mathbb{I}) P^T.
\end{equation}
Similarly, it can be shown that 
\begin{equation}
    U^R=P'(\mathbb{I} \otimes F) D(\tilde \Lambda) (\mathbb{I}    \otimes  F_d^{\dagger}) P'^T.
\end{equation}
where $P'=\sum_k X^k \otimes \ket{k}\bra{k}$ is the generalized {\sc cnot} gate having control on the second qudit and $D(\tilde \Lambda)=\sum_{a,b}\tilde \lambda_{a,b} \ket{a}\bra{a} \otimes  \ket{b}\bra{b}$, is diagonal matrix (not unitary in general!) obtained from $\tilde \Lambda=F_d \Lambda F_d$. 
{\blu
\section{Known algorithm for generating dual unitary and 2-unitary gates \label{app:algo}}
Iterative algorithms or maps for obtaining dual unitaries and perfect tensors were introduced in Ref.~\cite{SAA2020}. These maps are defined on the space of bipartite unitary operators, and one complete action of the map consists of two steps: (1) start with seed unitary $U_0$ and apply $R$ operation on $U_0$, to obtain $U_0^R$. (2) In general, $U_0^R$ is not unitary, take the nearest unitary $U_1$ to $U_0^R$ given by its polar decomposition, $U_1=U_0^R \left(\sqrt{U_0^{R\dagger} U_0^R}\right)^{-1}$. Repeating the same steps for $U_1$, for sufficiently large iterations, the algorithm converges to dual unitaries (up to numerical precision) with a high probability.
Starting from random unitaries, this algorithm also converges to perfect tensors in local dimension $d=3$ for some seed unitaries \cite{SAA2020}. For larger local dimensions, one of the ways to obtain obtain perfect tensors with a large probability is to take polar decomposition to $U_0^{R \Gamma}:=\left(U_0^{R}\right)^{\Gamma}$. The seed unitaries crucially determine whether the map converges to the perfect tensor or not, as was illustrated in Ref.~\cite{SRatherAME46} for local dimension $d=6$.

Starting with $U_0$ of the form given in Eq.~(\ref{eq:U_Diag_Max_Ent}) obtained from a unimodular vector containing random phases, the algorithms discussed in Ref.~\cite{SAA2020} converge to perfect tensors for several smaller local dimensions, including  $d=6$. For random phases, in most of the cases, the perfect tensors obtained from these numerical searches remain diagonal in the maximally entangled basis due to the invariance of the block-structure as discussed in the main text \footnote{Block structure is not guaranteed to be preserved if $U_0^R$ or $U_0^{\Gamma R}$ are not of full rank as polar decomposition is not uniquely defined in these cases.}. The corresponding phases satisfy conditions Eq.~(\ref{eq:lam_dual}) and Eq.~(\ref{eq:lam_T_dual}), and thus can be used to obtain biunimodular vectors for $F_d \otimes F_d$ that also satisfy Eq.~(\ref{eq:lam_T_dual}).

 \section{Distributions of numerically generated biunimodular vectors \label{app:numerics}}
Using random unimodular vectors as initial conditions for the iterative procedure for generating biunimodular vectors discussed in the main text, for sufficiently large number of iterations, we obtain unimodular vectors that are close to biunimodular vectors. The distribution of $\Delta(U^R):=\|U^RU^{R \dagger}-\mathbb{I}\|$ for bipartite unitary operators obtained from numerically generated biunimodular vectors is shown in the main part of Fig.(\ref{fig:biuni_conv}). As can be seen from the figure, the distribution is peaked around $\Delta(U^R)=0$ illustrating that given unitary operators are (close to) dual unitaries. For local dimension $d=3$, in some cases it turns out that $\Delta(U^{\Gamma}):=\|U^{\Gamma}U^{\Gamma \dagger}-\mathbb{I}\|=0$ resulting in perfect tensors. For local dimensions $d=4$ and $5$, numerical results (not shown here) show that one can obtain perfect tensors in these dimensions as well using the same iterative procedure. However, in local dimensions $d\geq6$, the minimum of $\Delta(U^{\Gamma})$ obtained was close to $3.92$ and hence did not result in perfect tensors. A way to obtain perfect tensors in local dimension $d=6$ is slightly modifying the iterative procedure for as discussed in the main text.

All biunimodular vectors are known in the two-qubit case $d^2=4$ therefore we focus on local dimensions $d>2$. We illustrate the convergence of the iterative map to biunimodular vectors (up to a numerical precision) by showing that the bipartite unitaries obtained from these, using Eq.~(\ref{eq:U_Diag_Max_Ent}), are close to dual unitaries i.e., $U^RU^{R\dagger} \approx \mathbb{I}$. We generate random unimodular vectors consisting of entries $e^{\theta_{ij}}$, where $\theta_{ij}$ are sampled uniformly in $[0,2 \pi)$. From these randomly generated unimodular vectors, we construct random bipartite unitary operators from Eq.~(\ref{eq:U_Diag_Max_Ent}). Using the Frobenius norm $||A||=\sqrt{\Tr A A^{\dagger}}$, distributions of the quantity $\Delta(U^R):=||U^RU^{R\dagger}-\mathbb{I}||$, for random unitaries are shown in the insets of Fig.~(\ref{fig:biuni_conv}); top part of the figure for local dimension $d=3$ and bottom one for $d=6$. As we are interested in perfect tensors, we also show the distribution of the quantity $\Delta(U^{\Gamma}):=||U^{\Gamma}U^{\Gamma \dagger}-\mathbb{I}||$ for given bipartite unitaries. For dual unitaries, $\Delta(U^R)=0$ and for perfect tensors, $\Delta(U^R)=\Delta(U^{\Gamma})=0$. As can be seen from the Figure insets, the probability of obtaining dual unitaries and perfect tensors from random sampling is very close to zero. The number of random realizations taken in each case is $10^4$.

\begin{figure*}
\centering
    \includegraphics[scale=0.6]{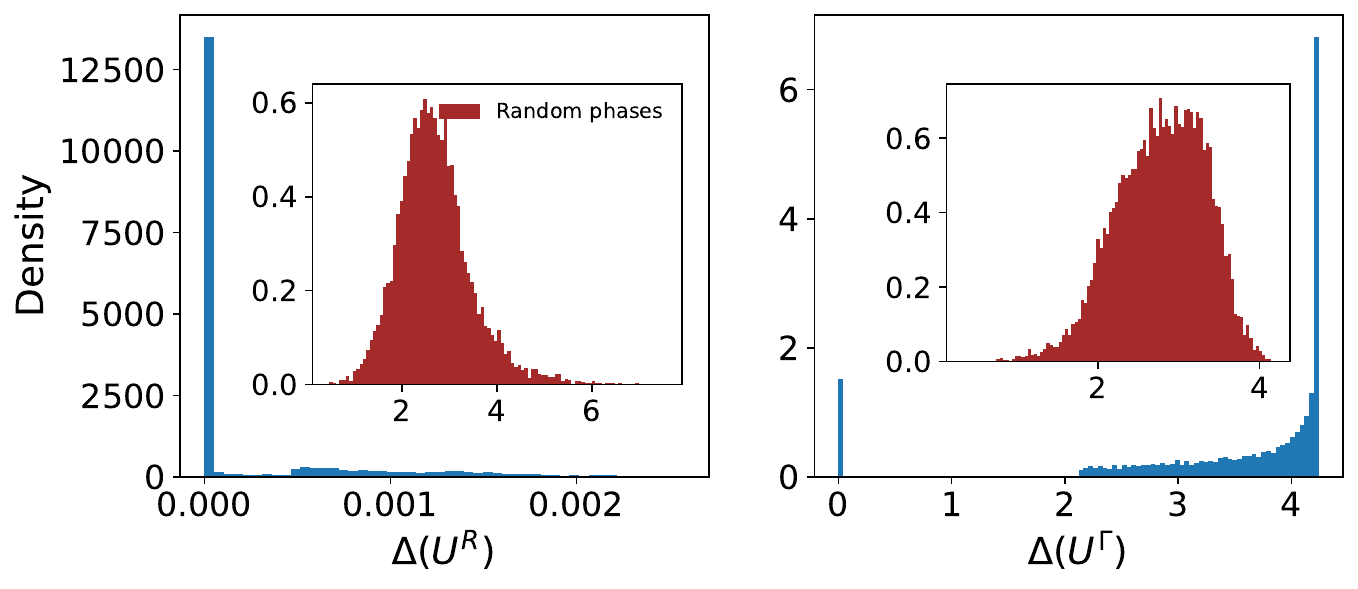}
    \includegraphics[scale=0.6]{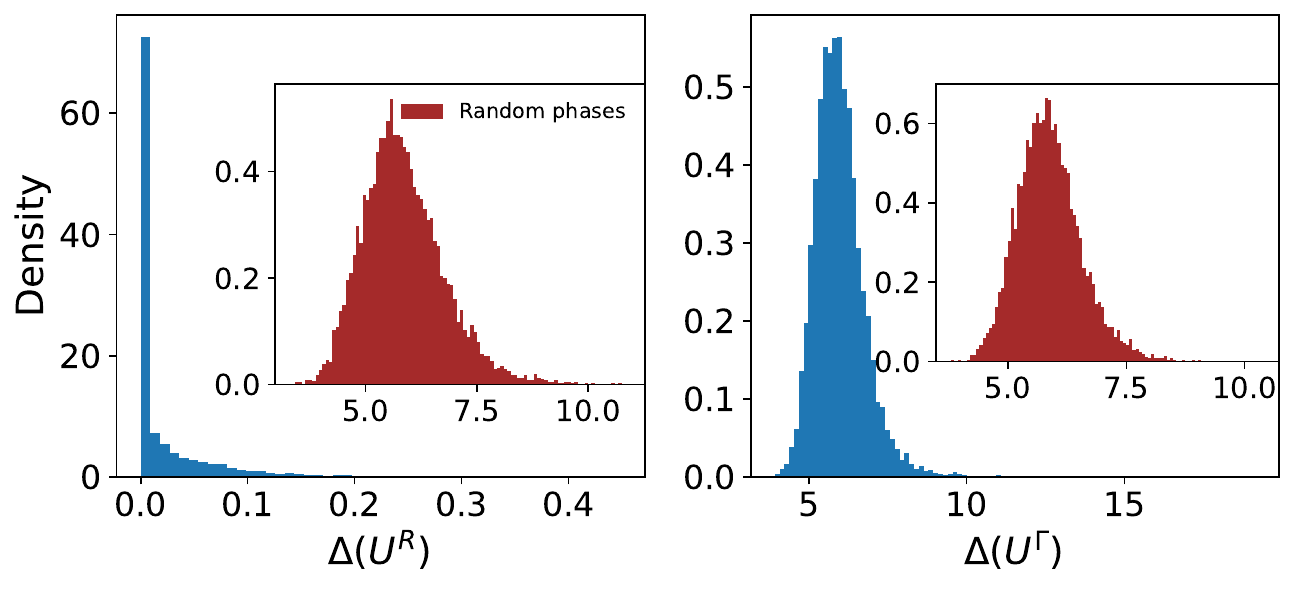}
    \caption{{\bf Top}: (Left) The (normalized) distribution of $\Delta(U^R):=||U^RU^{R\dagger}-\mathbb{I}||$ (where $||\cdots||$ is the Frobenius norm) is shown in the inset for random unitaries obtained from random unimodular vectors of length 9 consisting of entries $e^{i \theta_{ab}}$ with $\theta_{ab}$ sampled uniformly in $[0,2 \pi)$. The distribution of $\Delta(U^R)$ is shown in the main part of the figure for random unitaries of size $d^2=9$ obtained from biunimodular vectors resulting from the iterative procedure. The distribution is peaked around $\Delta(U^R)=0$ illustrating the convergence to dual unitaries.  (Right) The distribution of $\Delta(U^{\Gamma}):=||U^{\Gamma}U^{\Gamma \dagger}-\mathbb{I}||$ is shown in the inset for random unitaries obtained from random unimodular vectors. In the main part of the figure, the distribution of $\Delta(U^{\Gamma})$ is shown, and it is observed that the distribution has a small peak around $\Delta(U^{\Gamma})=0$ resulting in perfect tensors in $d^2=9$.
        {\bf  Bottom }: (Left) The distribution of $\Delta(U^R)$ is shown in the inset for random unitaries obtained from random unimodular vectors of length 36 consisting of entries $e^{\theta_{ab}}$ with $\theta_{ab}$ sampled uniformly in $[0,2 \pi)$. The distribution of $\Delta(U^R)$ is shown in the main part of the figure for random unitaries of size $d^2=36$ obtained from biunimodular vectors resulting from the iterative procedure. The distribution is peaked around $\Delta(U^R)=0$ illustrating the convergence to dual unitaries.  (Right) The distribution of $\Delta(U^{\Gamma}) $ is shown in the inset for random unitaries obtained from random unimodular vectors. In the main part of the figure, the distribution of $\Delta(U^{\Gamma})$ is shown and unlike the $d^2=9$, $\Delta(U^{\Gamma})=0$ is not obtained in this case. The number of realizations taken in each local dimension is $10^4$ and each realization of unimodular vector was iterated $10^4$ times.
    }
    \label{fig:biuni_conv}
\end{figure*}
}

\begin{figure*}
    \centering
    \includegraphics[scale=0.75]{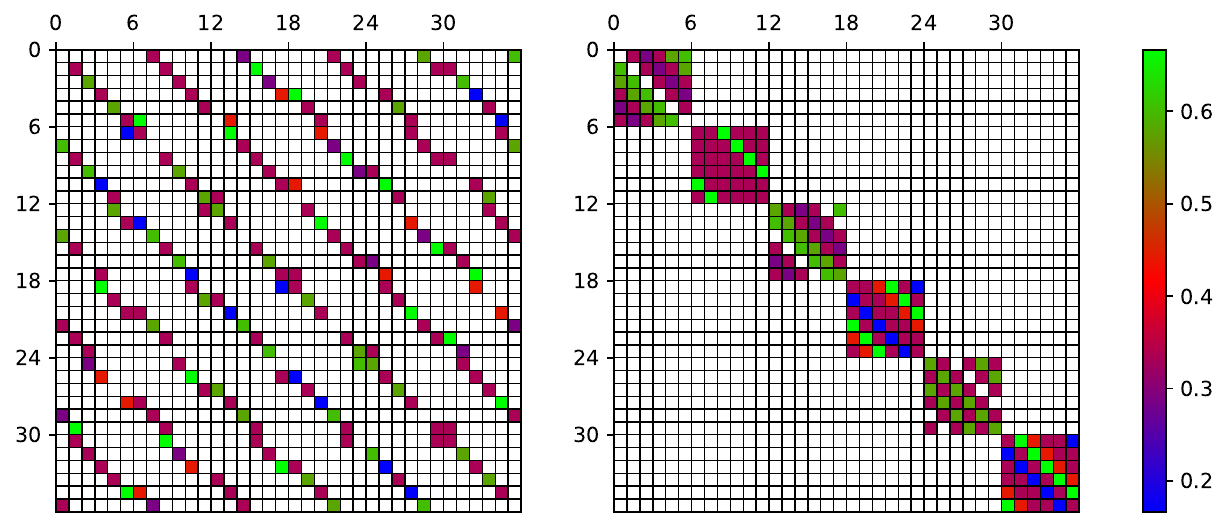}
    \includegraphics[scale=0.75]{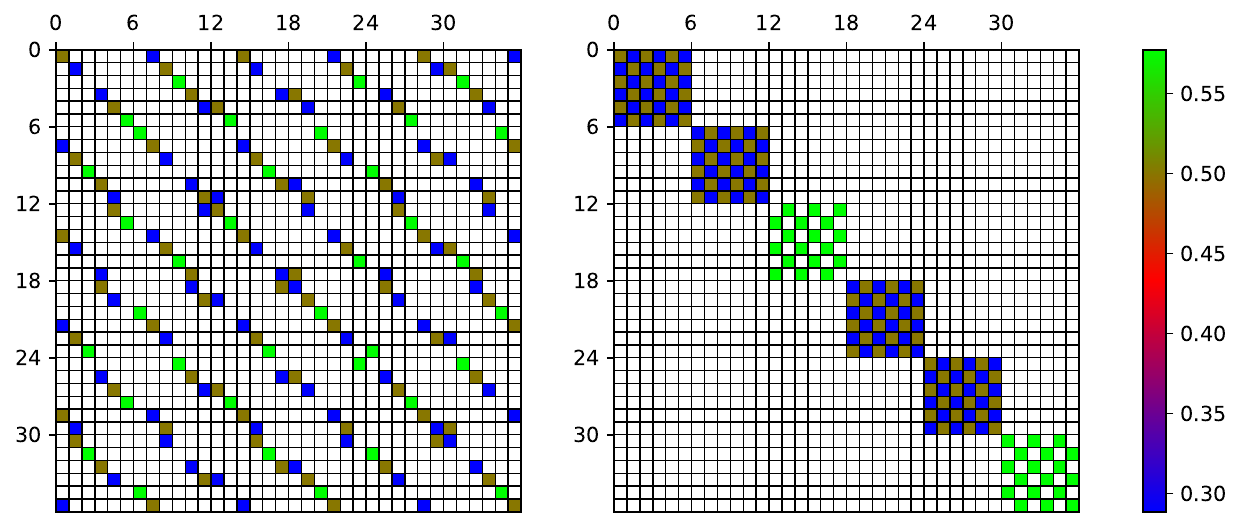}
    \includegraphics[scale=0.75]{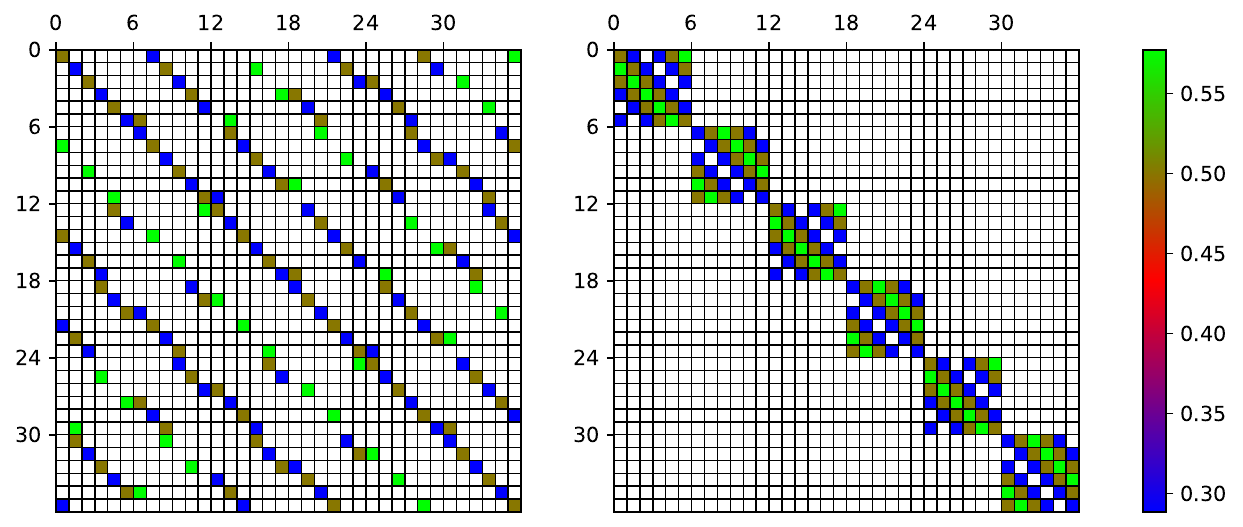}
    \caption{{\bf Top}: (Left) Absolute values of the non-zero entries of the perfect tensor $\mathcal{U}(\Lambda_1)$ obtained from the biunimodular vector $\Lambda_1$ are shown in the computational basis. (Right) 
    The block-structure determined by the non-zero entries consists of $6 \times 6$ unitary matrices each of size 6, and there are seven distinct non-zero absolute values.
    {\bf Middle}: (Left) Absolute values of the non-zero entries of the perfect tensor $\mathcal{U}(\Lambda_2)$ obtained from the biunimodular vector $\Lambda_2$ are shown in the computational basis. (Right) 
    The corresponding block-structure is shown, and there are only three non-zero distinct absolute values.
    {\bf Bottom}: (Left) Absolute values of the non-zero entries of the perfect tensor $\mathcal{U}(\Lambda_3)$ obtained from the biunimodular vector $\Lambda_3$ are shown in the computational basis. (Right) The corresponding block-structure is shown, and there are also only three non-zero distinct absolute values.
   }
    \label{fig:U_36_block}
\end{figure*}
 \end{document}